\documentclass{aa}
\usepackage[varg]{txfonts}
\usepackage{graphicx}

\begin{document}

\title{KKH~22, the first dwarf spheroidal satellite of IC~342}

\author{Igor D. Karachentsev\inst{1}
\and Lidia N. Makarova\inst{1}
\and R. Brent Tully\inst{2}
\and Gagandeep S. Anand\inst{2}
\and Luca Rizzi\inst{3}
\and Edward J. Shaya\inst{4}
\and Viktor L. Afanasiev\inst{1}}

\institute{Special Astrophysical Observatory, the Russian Academy of Sciences,Nizhnij Arkhyz,    Karachai-Cherkessian Republic, Russia 369167\\\email{ikar@sao.ru}
\and
Institute for Astronomy, University of Hawaii, 2680 Woodlawn Drive, Honolulu, HI 96822, USA
\and
W. M. Keck Observatory, 65-1120 Mamalahoa Hwy, Kamuela, HI 96743, USA
\and
Astronomy Department, University of Maryland, College Park, MD 20743, USA
}

\abstract {}
{
We present observations with the Advanced Camera for Surveys on the Hubble
Space Telescope of the nearby dwarf spheroidal galaxy KKH~22 = LEDA~2807114
in the vicinity of the massive spiral galaxy IC~342. 
}
{
We derived its distance of 3.12$\pm$0.19 Mpc using the tip of red giant branch 
(TRGB) method. We also used the 6 m BTA spectroscopy to measure a heliocentric 
radial velocity of the globular cluster in KKH~22 to be +30$\pm$10 km s$^{-1}$.
}
{
The dSph galaxy KKH~22 has the $V$-band absolute magnitude of --12$\fm$19 
and the central surface brightness $\mu_{v,0} = 24.1 ^m/\sq\arcsec$. 
Both the velocity and the distance of KKH~22 are consistent with the dSph 
galaxy being gravitationally bound to IC~342. Another nearby dIr galaxy, 
KKH~34, with a low heliocentric velocity of +106 km s$^{-1}$ has the TRGB 
distance of 7.28$\pm$0.36 Mpc residing in the background with respect to the
IC~342 group. KKH~34 has a surprisingly high negative peculiar velocity of
--236$\pm$26 km s$^{-1}$.
}
{}

\maketitle

\section{Introduction}

Nearby massive spiral galaxies with developed bulges (M~31, M~81, NGC~4258)
have many dwarf satellites. In contrast, the number of dwarf satellites
around spiral galaxies without apparent bulges (NGC~253, IC~342, M~101, NGC~6946)
is relatively small. This circumstance was noted by Ruiz et al. (2015) and
Javanmardi \& Kroupa (2020). The situation becomes even clearer if we consider
only dwarf spheroidal (dSph) satellites. Until recently, only one dSph
companion, SC~22 = LEDA~3097727, was known around NGC~253. In recent years,
Sand et al. (2014) and Toloba et al. (2016) discovered two new
dSph satellites near NGC~253: Scl-MM-Dw1 and Scl-MM-Dw2. For a long time, only late-type
dwarf satellites were known around M101. Now, Danieli et al. (2017),
Karachentsev \& Makarova (2019), and Bennet et al. (2019) have added the following
four dSph satellites to them: M101-Dw~A, M101-df~2, M101-df~3, and M101-Dw~9. However,
no dSph satellites have been discovered around the massive late-type spirals
IC~342, NGC~6946 or NGC~628.

IC~342 is the nearest massive late-type (Scd) spiral galaxy, which is situated in a zone of 
considerable extinction ( A$_B$ = 2.02 mag, Schlafly \& Finkbeiner, 2011) at a distance
of 3.28 Mpc (Saha et al. 2002). Its stellar mass corresponds to 10.60~dex in M$_\odot$,
which is comparable with the stellar mass of the Milky Way (10.78~dex) and M~31 (10.73~dex),
as well as the mass of NGC~253 (10.98~dex), M~101 (10.79~dex), and NGC~6946 (10.99~dex).

\begin{figure*}[h]  
\centering
 \includegraphics[height=14cm]{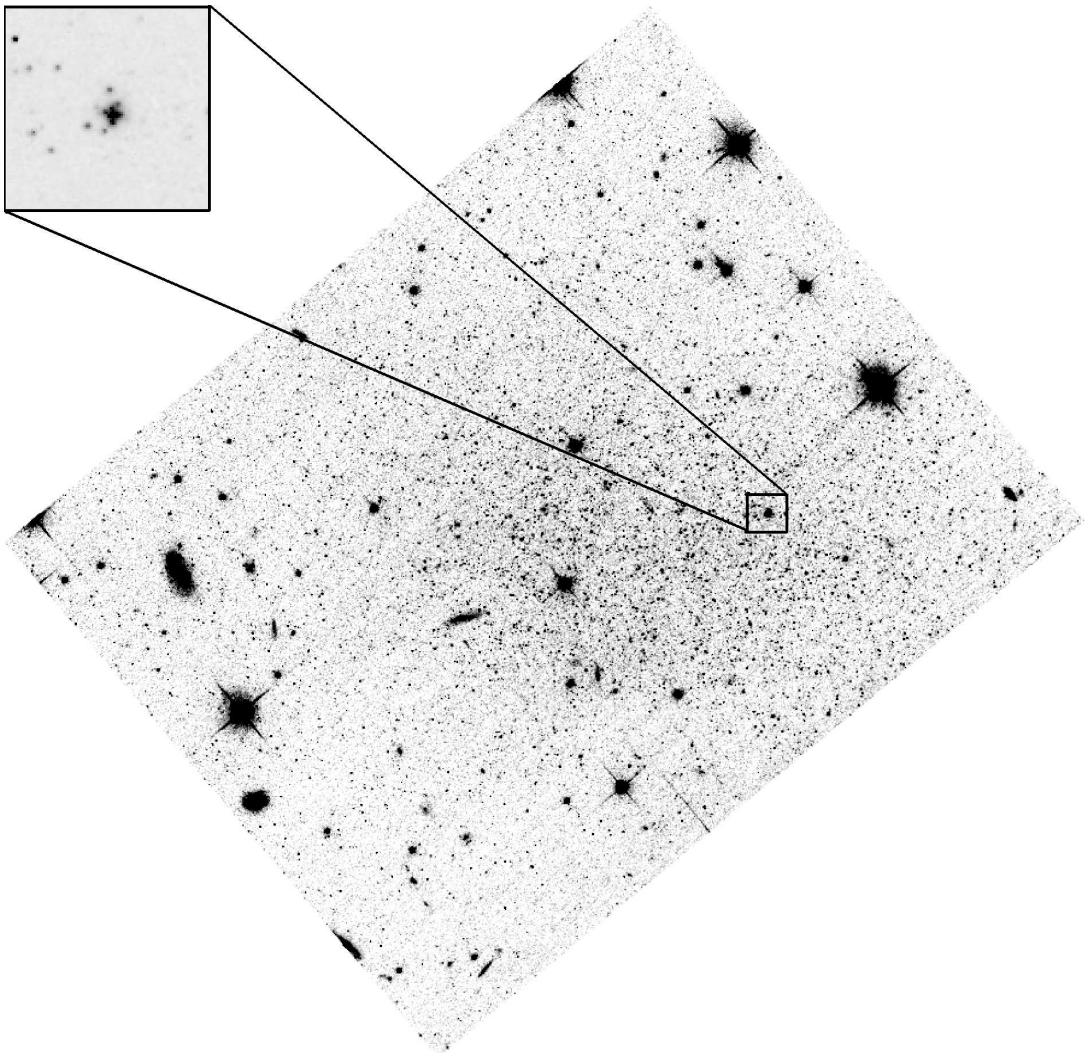}
 \caption{HST/ACS image of KKH~22 through the F814W filter. The image size is
        116 x 93 arcsec. North is up and east is left. The 5 arcsec region
        highlighted by the square is shown to contain a globular cluster.}
\label{fig:fig1}
\end{figure*} 

The first systematic attempt to search for dwarfs around IC~342 was undertaken by 
Borngen \& Karachentseva (1985). The authors used photographic plates which were obtained with
the wide-field Tautenburg 2 m telescope. Fifteen low surface brightness objects were 
found in the region of 40 square degrees around IC~342. However, subsequent
observations in the 21 cm line have shown that the radial velocities of these
galaxies are in the range of 830 -- 2500 km s$^{-1}$. Those candidate IC~342 
satellites turned out to be background dwarf galaxies.

At present, there are nine galaxies for which the IC~342 is the most significant
neighbour (the main disturber). All of them belong to late-type systems: 
Irr, Im, Sm, and Sd, which are listed in Section 5 (Table 2). Also, a dwarf irregular galaxy 
KKH~34 with a heliocentric velocity of V$_h$ = 105 km s$^{-1}$ and an apparent magnitude of
B$_T$ = 17.1 mag, but with an uncertain distance estimate, has been a candidate as a
remote companion to IC~342.

In this article, we report on measuring the distance and radial velocity of 
the galaxy KKH~22, which, by these parameters, has been revealed to be the first known dSph
satellite of the IC~342.  The object, KKH~34, that we observed turns out
to be a background dwarf galaxy. 

\section{HST observations and TRGB distance}

A northern (RA = 03:44:56.7, DEC = 72:03:52, J2000) low surface brightness
galaxy KKH~22 (LEDA~2807114) was found by Karachentsev et al. (2001). It was
not detected in the $HI$-line with the 100 m Effelsberg radio telescope at a
level of 8 mJy. The galaxy is also undetected in the H$\alpha$ line with an
upper limit of $5.7\times 10^{-16}$ erg cm$^{-2}$s$^{-1}$ (Kaisin \& Karachentsev 2013).
These properties indicate it to be a dwarf spheroidal galaxy.

Observations of KKH~22 were performed with the Advanced Camera for Surveys (ACS)
aboard the Hubble Space Telescope (HST) on October 14, 2019 as a part of SNAP project
15922 (PI R.B. Tully). Two exposures were made in a single orbit with the
filters F606W (760 s) and F814W (760 s). The F814W image of the galaxy is
presented in Fig.1. In the western part of KKH~22, which is highlighted by the $5\arcsec\times5\arcsec$
square, we found a globular cluster which is shown in the upper left side of Fig.1.
Photometry of the cluster yields its total magnitude $V = 21.42$ and the
colour $V-I = 0.86$ within the aperture 1$\farcs4$.

We used the ACS module of the DOLPHOT package
(http://purcell.as.arizona.edu/dolphot/) by Dolphin (2002) to perform photometry
of resolved stars based on the recommended recipe and parameters. Only stars
with good-quality photometry were included in the analysis. We selected the stars with a signal-to-noise ratio (S/N) of at least four in both filters,
and with DOLPHOT parameters $crowd_{F606W}+crowd_{F814W}\leq 0.8$,
$(sharp_{F606W}+sharp_{F814W})^2\leq 0.075$.
Artificial stars were inserted and recovered using the same reduction procedures to
accurately estimate photometric errors. The resulting colour-magnitude
diagram (CMD) in F606W -- F814W versus F814W is plotted in Fig.2.

\begin{figure}  
\centering
 \includegraphics[height=10cm]{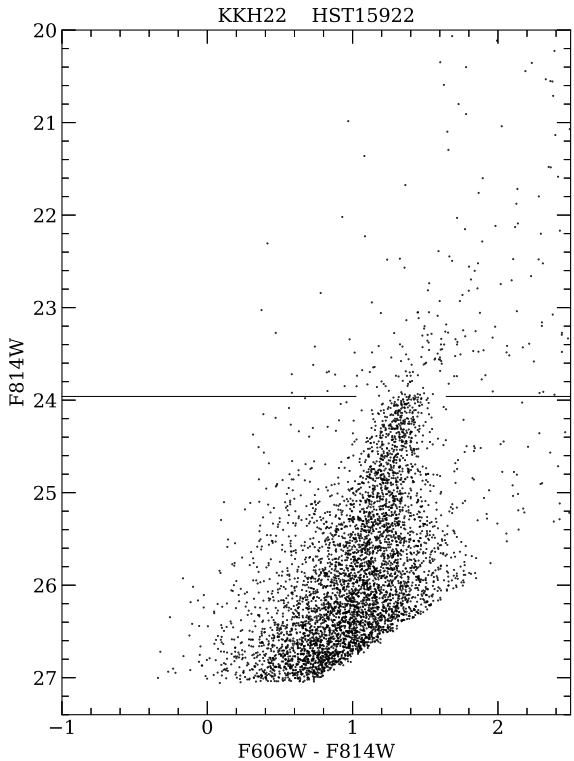}
 \caption{Colour-magnitude diagram of KKH~22. The TRGB position is indicated by
        the horizontal line.}
\label{fig:fig2}
\end{figure}  

A maximum-likelihood method (Makarov et al. 2006) was applied to estimate
the magnitude of the tip of the red giant branch (TRGB). We found
F814W(TRGB) to be 23$\fm96\pm0\fm13$. Following the zero-point calibration
of the absolute magnitude of the TRGB developed by Rizzi et al. (2007),
we obtained $M$(TRGB) = --4.09. Assuming  $E(B-V) = 0.340$ from Schlafly \&
Finkbeiner (2011) as for foreground reddening, we derived the true distance
modulus of $(m - M)_0 = 27.47\pm0.13$ or the distance $D = 3.12\pm0.19$ Mpc.

\section{Spectral observations}

The globular cluster of KKH~22 was observed in the long-slit mode of
the SCORPIO-2 focal reducer at the 6 m BTA telescope (Afanasiev \&
Moiseev 2011; Afanasiev et al. 2017), using the VPHG1200@860 grism as
a disperser. The length of the slit was 6$\arcmin$, while its width
was 2$\arcsec$. The spectral resolution was about 5\,\AA. During our 
observations, we used the technique of subtracting the sky background 
using the 'nod-shuffle' method  (Glasebrook \& Bland-Hawthorn 2001). To do 
this, we obtained a series of exposures lasting 900 seconds in which the object 
was sequentially shifted along the slit by a gap of +-8$\arcsec$ relative 
to the centre.  A total of five pairs of such expositions were received. 
The average seeing was 1$\farcs$8. The slit position, passing through 
several nearby foreground stars, is shown in Fig. 3.

\begin{figure*}[h]  
\centering
\includegraphics[height=8cm]{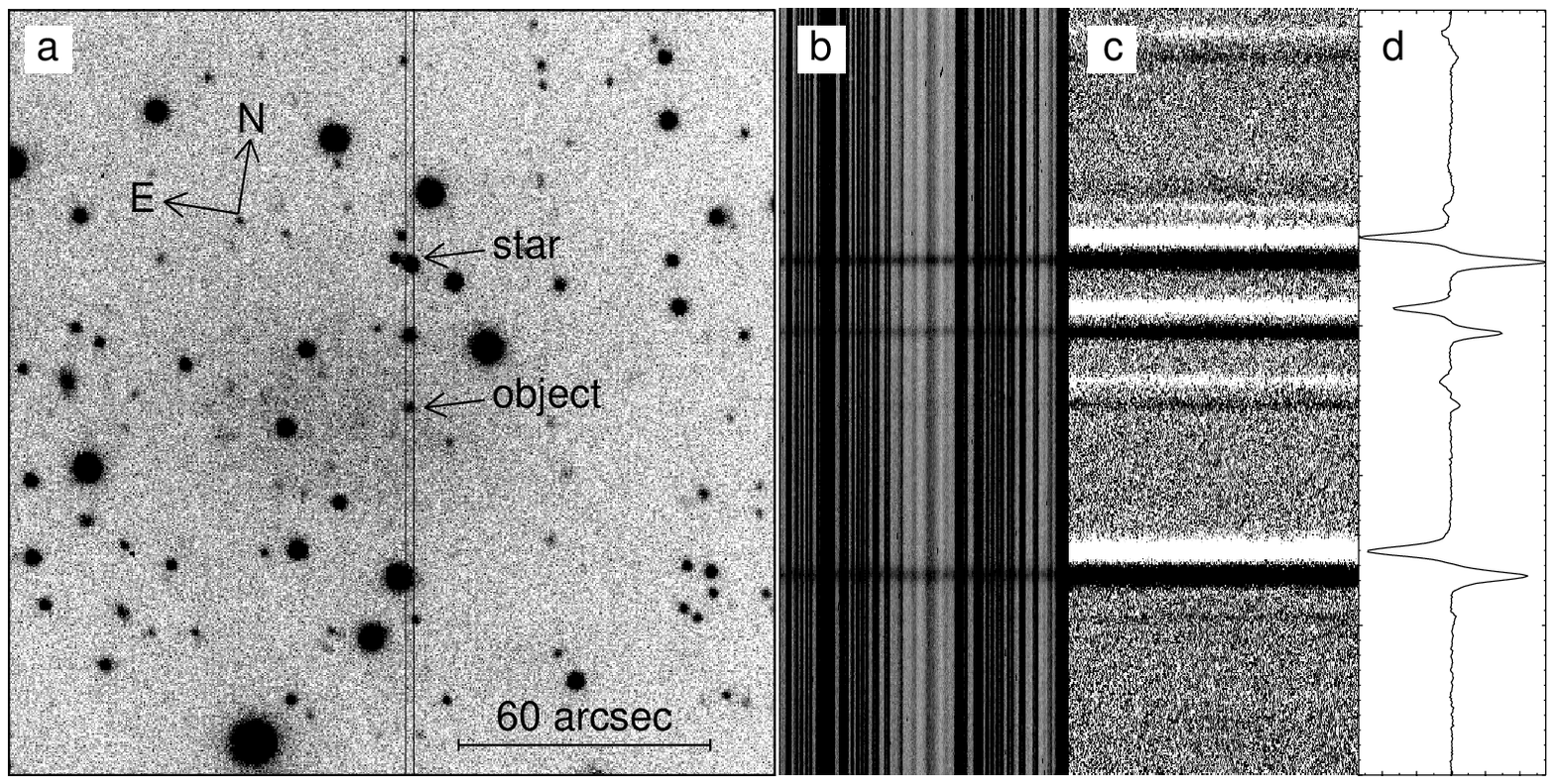}
\caption{ Identification map and spectra of KKH22. (a) Localisation of 
the slit on the  KH22  globular cluster, (b) the raw spectrum in the area 
of the infrared calcium triplet without subtracting the sky background, (c) the result of subtracting the sky background by the nod-shuffle procedure, 
and (d) the cross-section 2D spectra  along the slit after the sky subtraction.}
\label{fig:fig3}
\end{figure*}

\begin{figure*}[h] 
\centering
\includegraphics[height=10cm]{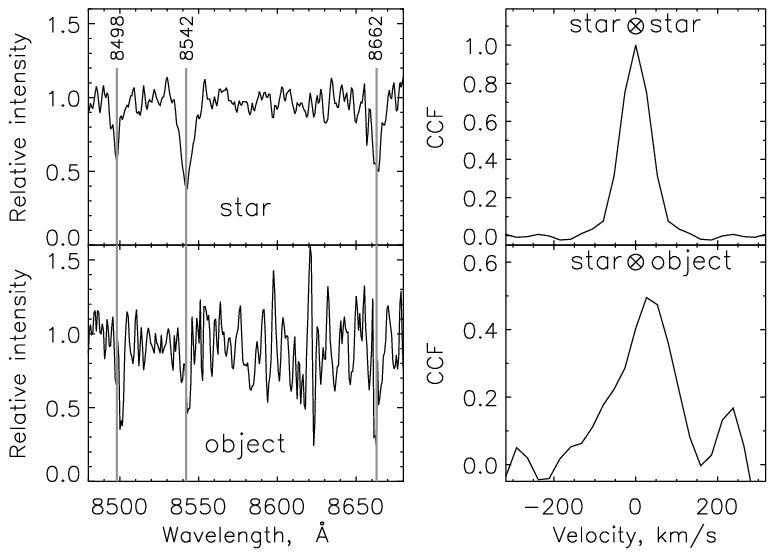}
\caption{Spectra showing the calcium triplet 8498/8542/8662\AA~ lines for the
        globular cluster and for the neighbouring foreground star (left-hand panel) 
and cross-correlation functions showing that the observed radial velocity of 
the object is 36$\pm$10 km/s (right-hand panel).}
\label{fig:fig4}
\end{figure*}

 The data reduction was carried out in a standard way using the IDL-based
software package for reducing long-slit spectroscopic data obtained with SCORPIO-2.
The data reduction included the following steps: bias subtraction, line
curvature and flat-field corrections, linearisation, and night sky 
subtraction. In the last step, unlike the standard one, the nod-shuffle 
algorithm was used. It consisted in the fact that two images were successively 
subtracted, in which the object was shifted by 16$\arcsec$ along the slit. 
The result of this subtraction is shown in panel (C) of Figure 3. Then the 
five pairs of images obtained in this way were added together, and the resulting 
spectrum of the object and the reference star was obtained by integrating the 
image modularly in a 20" wide strobe along the slit (see left panel of Fig. 4)

Spectra of the target as well  as of the reference  star  falling
in the slit are shown in the left-hand panel of Fig.4. Vertical lines indicate
the positions of the calcium triplet lines: 8498\AA , 8542\AA , and 
8662\AA. Using the cross-correlation method, we measured the observed 
radial velocity of $+36\pm$10 km s$^{-1}$, which gives the heliocentric radial 
velocity of the globular cluster to be +30 km s$^{-1}$. This velocity value 
is consistent with KKH~22 being gravitationally bound to IC~342, which has 
$V_{hel} = +29\pm1$ km s$^{-1}$ (Crosthwaite et al. 2000).

\section{Basic properties of KKH~22}

In the Updated Nearby Galaxy Catalogue (UNGC, Karachentsev et al. 2013)
KKH~22 was classified as a transition (Tr) type dwarf system. According
to UNGC, the linear Holmberg's diameter of the galaxy is 1.32 kpc, and
the stellar mass is $(M_*/M_{\odot}) = 6.81$ dex at the measured distance of
3.12 Mpc. The absence of a noticeable flux from KKH~22 in the $HI$-line
(Karachentsev et al. 2001) and in the $H\alpha$ line (Kaisin \& Karachentsev
2013), as well as in the FUV-band indicates that the galaxy belongs to the class of spheroidal
dwarfs with old stellar populations. The distribution of stars on
the CM-diagram (Fig.2) is concordant with this statement.

Some basic characteristics of the dwarf galaxy KKH~22 are presented in
Table 1. We performed surface photometry of the galaxy and determined
its integrated magnitude $V= 16\fm34\pm0\fm12$ and colour index $V-I =
1.31\pm0.09$. Taking into account Galactic extinction (Schlafly \& Finkbeiner
2011), the integrated apparent magnitude and integrated colour are
$V_0 = 15\fm28$ and $(V-I)_0 = 0.83$. The central surface brightness of
the galaxy, $\mu_{v,0} = 24.1\pm0.2^m/\sq\arcsec$, and its absolute
magnitude, $M_{v,0} = -12\fm19$, are typical of dwarf spheroidal
satellites around nearby massive spirals such as M~31 and M~81.

\begin{table}
 \caption{Properties of KKH~22.}
 \begin{tabular}{ll} \\ \hline

 Parameter         &           Value     \\
\hline
 RA (J2000)            &             03:44:56.6   \\
 DEC (J2000)             &          +72:03:52      \\
$(m-M)_o$, mag              &       27.47$\pm$0.13\\
 $D$, Mpc                 &          3.12$\pm$0.19\\
 $V_{hel}$, km s$^{-1}$    &                30$\pm$10\\
 $V_{LG}$, km s$^{-1}$     &                251\\
 $E(B-V)$               &            0.340\\
 $E(V-I)$                  &         0.477\\
 $V_0$, mag                  &       15.28$\pm$0.12 \\
 $(V-I)_o$                      &    0.83$\pm$0.09\\
 $\mu_{V,0}$, mag/arcsec$^{-2}$  &   24.1$\pm$0.2\\
 $\mu_{I,0}$, mag/arcsec$^{-2}$  &   23.3$\pm$0.2\\
 $M(V)_0$                     &    --12.19 \\
 $A_{26}$, kpc                &      1.32    \\
 axial ratio                   &     0.52    \\
 $\log(M_*), M_{\odot}$       &      6.81    \\
 $\log(M_{HI}), M_{\odot}$     &    $<$6.34  \\
 $\log[SFR(H_{\alpha})], M_{\odot}$/yr       &     $<-$4.92  \\
 $\log[SFR(FUV)], M_{\odot}$/yr       &     $<-$4.15  \\
\hline
 Globular cluster: &  \\
 RA (J2000)                &         03:44:50.49 \\
 DEC (J2000)                 &      +72:03:56.4\\
 $V$, mag                    &       21.42$\pm$0.03\\
 $V-I$                        &      0.86$\pm$0.03\\
 $M_{v,0}$ , mag               &     --7.11 \\
 $(V-I)_0$                     &     0.38\\
\hline
\end{tabular}
\end{table}

Figure 5 presents the distribution of the most luminous dSph satellites 
of the Milky Way (Fornax, Leo~I, Sculptor, and Leo~II), M~31 (NGC~147, And~II, 
And~I, CasSph, and PegSph), M~81 (KDG61, KDG64, KDG63, and F8D1), and NGC~253 (SC~22) according 
to their integrated absolute B-magnitudes and linear Holmberg's diameters 
A$_{26}$, taken from UNGC. As one can see, the position of KKH~22 in this 
diagram, which is shown by an asterisk, does not stand out among other objects.

\begin{figure*}  
\centering
\includegraphics[height=9cm]{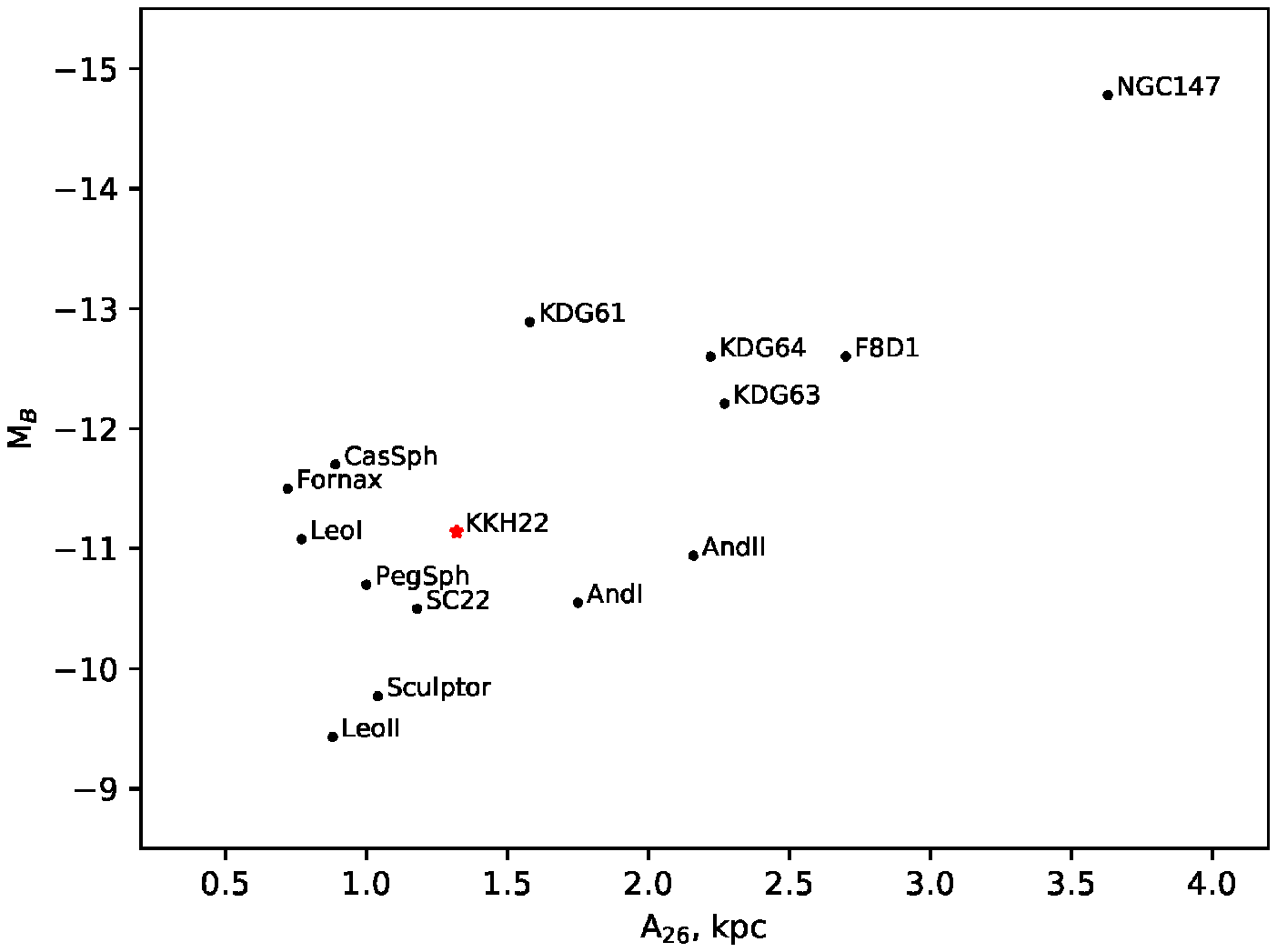}
\caption{Distribution of the most luminous dSph satellites 
of the Milky Way (Fornax, Leo~I, Sculptor, and Leo~II), M~31 (NGC~147, And~II, 
And~I, CasSph, and PegSph), M~81 (KDG61, KDG64, KDG63, and F8D1), and NGC~253 (SC~22) according 
to their integrated absolute B-magnitude and the linear Holmberg's diameter 
A$_{26}$.}
\label{fig:fig5}
\end{figure*}

The last rows in Table 1 contain integral characteristics of the globular
cluster. Given its absolute magnitude of $M_{v,0} = -7\fm11$, the
luminosity of the globular cluster contributes 1\% of the total luminosity
of the dwarf galaxy.

\section{KKH~22 as a member of the IC~342 group}

Judging by the distances and radial velocities of nearby galaxies, the gravitational
dominance zone of IC~342 includes the nine dwarf satellites listed in Table 2.
The table columns contain: (1) the galaxy name; (2,3) Galactic coordinates;
(4,5) the angular and linear projected separation from IC~342; (6) the radial velocity
of the expected satellite relative to IC~342; (7) the galaxy distance from the observer
measured via the TRGB; and (8,9) a dimensionless 'tidal index'
$\Theta_1$, indicating a density contrast which was contributed by the most significant 
neighbour ('main disturber') of the galaxy as well as the main disturber's name. 
The cases with positive $\Theta_1$ can be considered as objects bound to 
the main disturber. As seen, all nine dwarfs, besides KKH~34, are gravitationally
linked to IC~342. An updated version of the UNGC (http://www.sao.ru/lv/lvgdb)
contains links to the sources of the data that are used.

\begin{table*}
\caption{Known satellites of IC342.}
\begin{tabular}{lccrrrcrc} \\ \hline
 Galaxy  &    $l$   &   $b$   &  $r_p$ &  $R_p$ &    $\Delta V$   &  $D$ & $\Theta_1$ & Main disturber \\
         &   deg  &  deg  &  deg  & kpc &   km s $^{-1}$  &     Mpc  &  & \\
\hline
 IC342   & 138.17 & 10.58 & 0.00 &   0  &   0$\pm$1  &  3.28$\pm$0.30  & --0.1 & NGC 1569 \\
 KK35    & 138.20 & 10.31 & 0.27 &  16  & --95$\pm$3  &  3.16$\pm$0.32 & 2.4   & IC 342 \\
 UGCA86  & 139.76 & 10.65 & 1.59 &  91  &  36$\pm$5  &  2.98$\pm$0.28  & 1.1   & IC 342 \\
 KKH22   & 135.50 & 13.57 & 4.01 & 229  &   7$\pm$10 &  3.12$\pm$0.19  & 1.6   & IC 342 \\
 NGC1560 & 138.37 & 16.02 & 5.44 & 312  & --74$\pm$1  &  2.99$\pm$0.21 & 0.8   & IC 342 \\
 NGC1569  &143.68 & 11.24 & 5.55 & 318  &--138$\pm$3  &  3.19$\pm$0.10 & 1.1   & IC 342 \\
 Cam A    &137.25 & 16.20  &5.69 & 326  & --88$\pm$2  &  3.56$\pm$0.23 & 0.7   & IC 342 \\
 Cam B    &143.38 & 14.42 & 6.47 & 370  &  23$\pm$2  &  3.50$\pm$0.28  & 0.7   & IC 342 \\
 UGCA92   &144.71 & 10.52  &6.54 & 374  &--151$\pm$2  &  3.22$\pm$0.11 & 1.8   & NGC 1569 \\
 UGCA105  &148.52 & 13.66 &10.80 & 618  &  37$\pm$5  &  3.39$\pm$0.35  & 0.3   & IC 342 \\
\hline
 KKH34   & 140.42 & 22.35 &11.98  &687  &  51$\pm$2   & 7.28$\pm$0.36  & --0.6 & Maffei 2 \\
\hline
\end{tabular}
\end{table*}

 The average distance of the expected satellites from the observer is
3.23$\pm$0.09 Mpc, which practically coincides with the distance of the principal
galaxy IC~342. The average radial velocity difference of the identified
satellites, --49$\pm$25 km s$^{-1}$, shows a weak asymmetry towards low radial
velocities for the satellites. The distribution of galaxies in the IC~342 group
in Galactic coordinates is characterised by significant asymmetry,
with almost all of the dwarf companions residing at higher Galactic latitudes than
IC~342. Such an eccentricity of the IC~342 position can be attributed to
the substantial extinction in the Zone of Avoidance. 
Satellites within the influence of IC~342 that are closer to the Galactic plane may be
hidden by obscuration.
 The search for such objects by their emission in the 21 cm
line is a difficult task since their radial velocities fall within the velocity
range of the local Galactic hydrogen.

  The only dSph satellite, KKH~22 (at $b=13.6$ and with $A_B=1.7$), is at a projected separation of 229
kpc from IC~342. Dwarf spheroidal companions to the Milky Way, Leo~I and
Leo~II, have approximately the same spatial separations (250 kpc and 210 kpc).
These values are all close to the virial radii of IC~342 and the Milky Way.

\section{KKH~34}

 At a higher Galactic latitude, there is a dwarf irregular galaxy KKH~34=
PGC~095594, with a radial velocity close to the IC~342 velocity. Based on
a shallow CM-diagram, Karachentsev et al. (2003) determined the galaxy
distance to be 4.61 Mpc. Since this distance estimation was considered to be unreliable, KKH~34 was
included in our ongoing SNAP survey. An $I$-band image from a new ACS observation is
shown in the upper panel of Fig.6. The red giant branch with a TRGB position of
$I_{TIP} = 25\fm60$ is visible in the corresponding colour-magnitude diagram (bottom
panel of Fig.6). Taking the Galactic extinction of $A_I = 0\fm361$ into account,
we obtain $(m-M)_o = 29.31\pm0.11$ or
$D = 7.28\pm0.36$ Mpc for the galaxy distance module. Parameters for this galaxy are presented in the last
row of Table 2. 

\begin{figure*}  
\centering
\includegraphics[height=12cm]{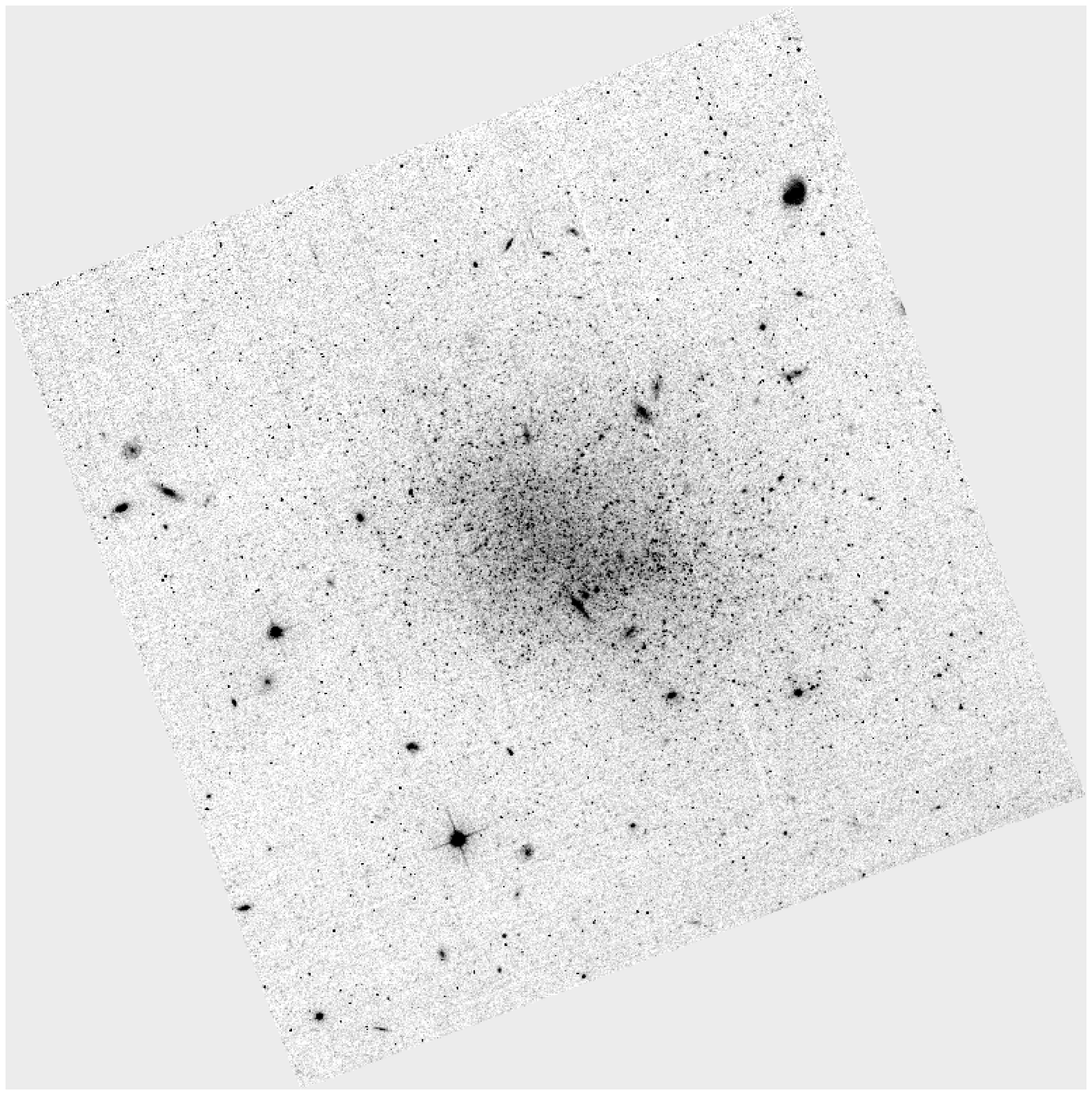}
\includegraphics[height=10cm]{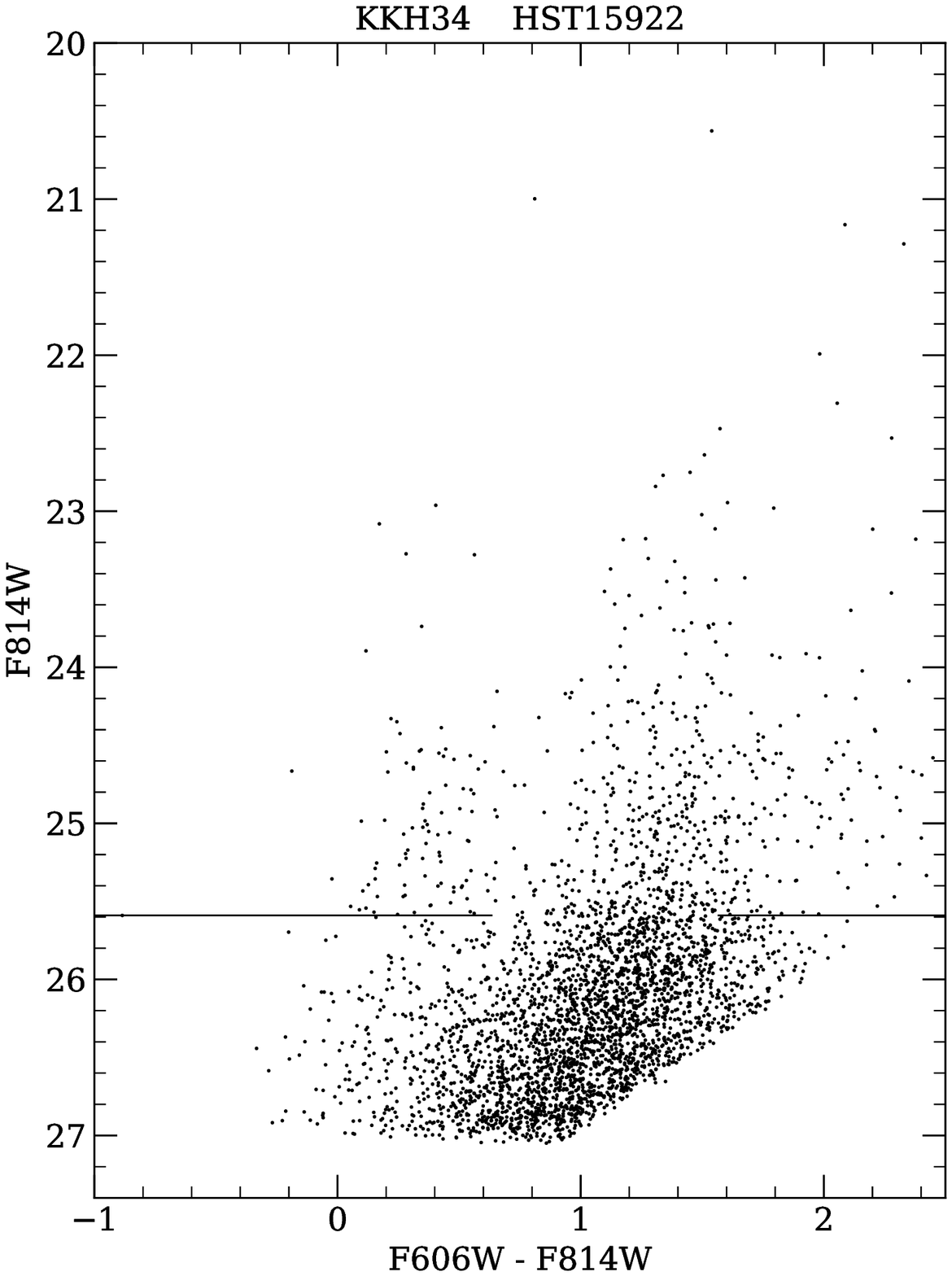}
\caption{ACS F814W-filter image of 1.5 x 1.5 arcmin (upper panel) and CM diagram
        (lower panel) for the dwarf irregular galaxy KKH~34.}
\label{fig:fig5}
\end{figure*}

At this revised distance, KKH~34 is 4~Mpc behind the IC~342 group, but only 2~Mpc from the Maffei group
(Anand et al. 2019a). 
At the radial velocity of KKH~34 relative to the Local Group
centroid of $V_{LG} = 295$ km s$^{-1}$ (Begum et al.  2008) and an assumed Hubble parameter of
73 km s $^{-1}$ Mpc$^{-1}$, the peculiar velocity of this galaxy is $-236\pm$26~km~s$^{-1}$. 
This motion towards us has the same sign but is greater than the $-128\pm33$~km~s$^{-1}$ of the 
Maffei group, which was reported by Anand et al.  The study of the Local Void from Tully et al. (2019) provides an
explanation for these motions.   The Local Void is generating a currently well-documented displacement of the Local Sheet
(Anand et al. 2019b) towards the negative supergalactic pole (towards negative SGZ).  With a realisation
of the full extent of the Local Void, it becomes apparent that the Local Void not only dominates at positive SGZ
in our proximity but also at positive SGX in the space between us and the Perseus-Pisces filament 
(Haynes \& Giovanelli 1986).  The Maffei group and KKH~34 somewhat further back are experiencing 
the push of this void.

\section{Concluding remark}

In the Introduction, we address the fact that luminous bulgeless galaxies suffer from a lack
of dwarf spheroidal satellites. Recently, Karachentsev \& Karachentseva (2019)
noted another feature of nearby massive Sc-Sd galaxies. Their mean estimate of
total mass via the orbital motions of small companions, taken in relation to the stellar
mass, is $<M_{orb}/M_*> = 22\pm5$. This quantity is two times and four times less than the
mean ratios for Sab-Sbc and for E, S0, and Sa galaxies, respectively. It is quite possible
that both of these features of the bulgeless galaxies have a common evolutionary
cause.
        
\begin{acknowledgements}
This work is based on observations made with the NASA/ESA Hubble Space
Telescope and the 6 metre BTA telescope. STScI is operated by the
Association of Universities for Research in Astronomy, Inc. under NASA
contract NAS 5--26555. The work in Russia is supported by RFBR grant
18--02--00005.
\end{acknowledgements}

\bigskip
{\bf  REFERENCES}
\bigskip

Afanasiev, V.L., Amirkhanian, V.R., Moiseev, A.V., et al. 2017, AstrBu, 72, 458

Afanasiev, V., \& Moiseev, A. 2011, Baltic Astronomy, 20, 363

Anand, G.S., Tully, R.B., Rizzi, L., et al. 2019a, ApJ, 872, L4

Anand, G.S., Tully, R.B., Rizzi, L., et al. 2019b, ApJ, 880, 52

Begum, A., Chengalur, J.N., Karachentsev, I.D., et al. 2008, MNRAS, 386, 1667

Bennet, P., Sand, D.J., Crnoevic, D., et al. 2019, ApJ, 885, 153

Borngen, F., \& Karachentseva, V.E., 1985, Astron.Nachr, 306, 301

Crosthwaite, L.P., Turner, J.L., \& Ho, P.T.P., 2000, AJ, 119, 1720

Danieli, S., van Dokkum, P., Merritt, A., et al. 2017, ApJ, 837, 136

Dolphin, A.E. 2002, MNRAS, 332, 91

Glasebrook, K., \& Bland-Hawthorn, J. 2001, PASP, 113, 107

Haynes, M.P., \& Giovanelli, R. 1986, ApJL, 306, L55

Javanmardi, B., \& Kroupa, P. 2020, arXiv:2001.02680

Kaisin, S.S., \& Karachentsev, I.D. 2013, AstrBu, 68,381

Karachentsev, I.D., \& Makarova, L.N. 2019, Astrophysics, 62, 293

Karachentsev, I.D., \& Karachentseva, V.E. 2019, MNRAS, 486, 3697

Karachentsev, I.D., Makarov, D.I., \& Kaisina, E.I. 2013, AJ, 145, 101

Karachentsev, I. D. Makarov, D.I., Sharina, M.E., et al. 2003, A\&A, 398, 479

Karachentsev, I.D., Karachentseva, V.E., \& Huchtmeier, W.K. 2001, A\&A, 366, 428

Makarov, D., Makarova, L., Rizzi, L., et al. 2006, AJ, 132, 2729

Rizzi, L., Tully, R.B., Makarov, D., et al. 2007, ApJ, 661, 815

Ruiz, P., Trujillo, I., \& Marmol-Queralto, E. 2015, MNRAS, 454, 1605

Saha, A., Claver, J., \& Hoessel, J.G. 2002, AJ, 124, 839

Sand, D.J., Crnoevic, D., Strader, J., et al. 2014, ApJ, 793L, 7

Schlafly, E.F., \& Finkbeiner, D.P. 2011, ApJ, 737, 103

Toloba, E., Sand, D.J., Spekkens, K., et al. 2016, ApJ, 816, L5

Tully, R.B., Pomar\`ede, D., Graziani, R., et al. 2019, ApJ, 880, 24

\end{document}